\documentclass{iopart}

\usepackage{graphicx}
\usepackage{calrsfs}
\usepackage{xcolor}
\usepackage{hyperref}

\begin{document}

\title{Skyrmionium Dynamics and Stability on One Dimensional Anisotropy Patterns}

\author{J. C. Bellizotti Souza$^1$,
  N. P. Vizarim$^2$, 
  C. J. O. Reichhardt$^3$, 
  C. Reichhardt$^3$,
  and P. A. Venegas$^4$}

\ead{jc.souza@unesp.br}

\address{$^1$ POSMAT - Programa de P\'os-Gradua\c{c}\~ao em Ci\^encia e Tecnologia de Materiais, S\~ao Paulo State University (UNESP), School of Sciences, Bauru 17033-360, SP, Brazil}

\address{$^2$ “Gleb Wataghin” Institute of Physics, University of Campinas, 13083-859 Campinas, S\~ao Paulo, Brazil}

\address{$^3$ Theoretical Division and Center for Nonlinear Studies, Los Alamos National Laboratory, Los Alamos, New Mexico 87545, USA}

\address{$^4$ Department of Physics, S\~ao Paulo State University (UNESP), School of Sciences, Bauru 17033-360, SP, Brazil}

\date{\today}

\begin{abstract}
  We examine a skyrmionium driven over a periodic anisotropy pattern, which consists of disorder free regions and disordered regions. For small defect densities, the skyrmionium flows for an extended range of currents, and there is a critical current above which it transforms into a skyrmion. For increased amounts of quenched disorder, the current needed for the skyrmionium to transform into a skyrmion decreases, and there is a critical disorder density above which a moving skyrmionium is not stable. In the moving state, the skyrmionium to skyrmion transformation leads to a drop in the velocity and the onset of a finite skyrmion Hall angle. We also find a reentrance effect in which the pinned skyrmionium transforms into a skyrmion just above depinning, restabilizes into skyrmionium at larger drives, and becomes unstable again at large currents. We also show that adding a transverse shaking drive can increase the lifetime of a moving skyrmionium by reducing the effect of the pinning in the direction of the drive.
\end{abstract}

\maketitle

\section{Introduction}

Magnetic textures such as skyrmions, skyrmionium, and merons have been attracting growing attention in basic science since they
represent a new class of particle like states that can have interesting static
and dynamical behaviors in the presence of
a driving force and quenched disorder \cite{Nagaosa13,Lin13,Woo16,EverschorSitte18,Gobel21a,Jani21,Reichhardt22a}.
They are also promising candidates for a number of applications
for memory \cite{Fert13,Wang19,Vakili21} and
novel computing architectures \cite{Zazvorka19,Back20,Zhang20a}.
For many of these applications, it is necessary
to understand how stable these textures are against
quenched disorder and driving forces
\cite{iwasaki_universal_2013, Reichhardt15, Jiang17,Litzius17,Juge21}.
Topological textures can be defined by their topological number $Q$,
or how many times their spin degrees of freedom can be wrapped around a sphere. When this topological number is one, $Q=1$, the state is a skyrmion and can exhibit a skyrmion Hall angle when moving. It is also possible to have a $Q=0$ texture that is called skyrmionium  \cite{Bogdanov99,Finazzi13, Zhang16e,Kolesnikov18,Zhang18d,Ishida20,Xia20a}. For applications, skyrmioniums have several
advantages over skyrmions in that their skyrmion Hall angle is zero,
making it possible for skyrmioniums to travel along narrow channels
without being pushed to the channel edge and annihilated due to the
skyrmion Hall effect
\cite{Kolesnikov18,Ishida20,Souza24}. Additionally,
skyrmioniums move twice as fast as skyrmions \cite{Kolesnikov18,Ishida20,Souza24}.
Recent work indicates, however, 
that skyrmionium is less stable under an applied current,
and it transforms into a skyrmion at higher drives \cite{Xia20a,Souza24}.
Skyrmioniums can also undergo
strong distortions with small Gilbert damping \cite{Ishida20}.
An open question is how stable skyrmionium is when
it is driven over quenched disorder, which may either
arise naturally in a sample or be introduced deliberately through
nanostructuring.

In this work, we consider atomistic simulations of a skyrmionium moving over a periodic quasi-one-dimensional array of high 
and low pinning regions,
where we vary the density of defects and the driving force.
For low defect densities, the skyrmionium can undergo stable motion
over an extended range of drives, and we find a critical current
above which the skyrmionium transforms into a skyrmion.
As the defect density increases,
the current at which the skyrmionium transitions into a skyrmion decreases.
The transformation to a skyrmion results in a drop
in the velocity as well as the appearance of a finite Hall angle.
Above
a critical disorder strength $\rho_{\rm def}^c$,
a moving skyrmionium is never stable.
There is also a large increase in the depinning threshold of the
texture above $\rho_{\rm def}^c$.
We observe a reentrant transition in which
the skyrmionium breaks up into a skyrmion
at drives just above depinning, restabilizes into a skyrmionium
at intermediate drives, and
becomes unstable again at higher drives.
The reentrance is produced by a
competition between the quenched disorder and the driving force,
since the driving force can partially reduce the effectiveness of the
pinning in the moving state.
We also show that applying a transverse ac drive to skyrmionium moving
over strong quenched disorder can increase the amount of time that the
skyrmionium remains stable.

\section{Methods}

\begin{figure*}
  \centering
  \includegraphics[width=\textwidth]{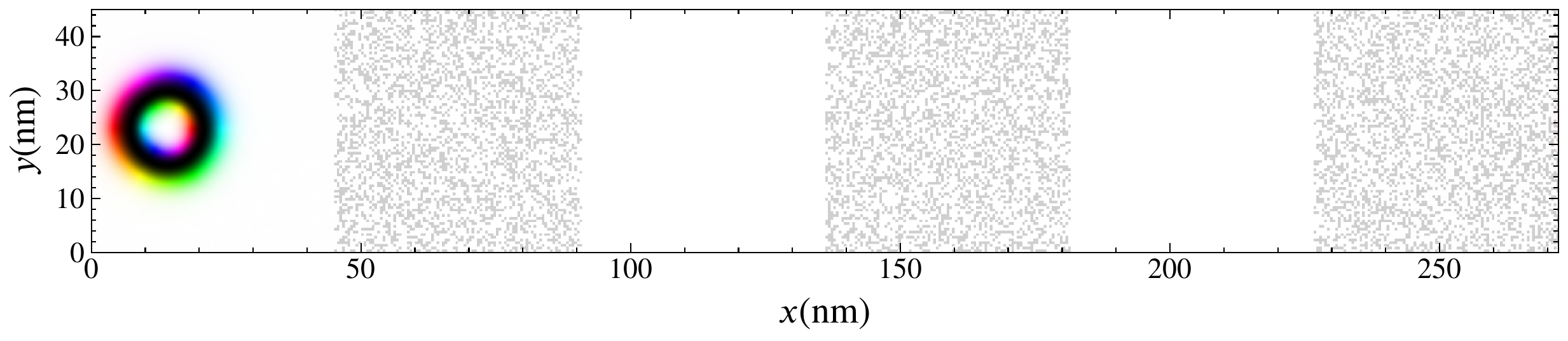}
  \caption{Image of the sample showing the periodic
    arrangement of defect regions.
    White regions are defect free, and the gray squares
    represent anisotropy defects in the pinned regions.
    The skyrmionium (rainbow colored ring) is initialized
    inside a clean region
    within the sample. Here $\rho_\mathrm{def}=17.4\%$.}
  \label{fig:1}
\end{figure*}

We use an atomistic model \cite{evans_atomistic_2018}
to model an ultrathin ferromagnetic film.
Our sample is of size 272 nm $\times$ 45 nm with periodic
boundary conditions along the $x$ and $y$ directions.
We apply a magnetic field perpendicular to the sample
along $+z$ at zero temperature, $T=0$ K.

The atomistic dynamics are governed by the Hamiltonian
\cite{evans_atomistic_2018, iwasaki_universal_2013, iwasaki_current-induced_2013}:

\begin{eqnarray}\label{eq:1}
  \mathcal{H}&=-\sum_{i, \langle i, j\rangle}J_{ij}\mathbf{m}_i\cdot\mathbf{m}_j
  -\sum_{i, \langle i, j\rangle}\mathbf{D}_{ij}\cdot\left(\mathbf{m}_i\times\mathbf{m}_j\right)
  -\sum_i\mu\mathbf{H}\cdot\mathbf{m}_i\\\nonumber
  &-\sum_{i \notin N} K_0\left(\mathbf{m}_i\cdot\hat{\mathbf{z}}\right)^2
  -\sum_{i \in N} K_d\left(\mathbf{m}_i\cdot\hat{\mathbf{z}}\right)^2 \ .
\end{eqnarray}

The ultrathin film is modeled as a square arrangement of atoms
with a lattice constant $a=0.5$ nm.
The first term on the right hand side is the exchange interaction
with an exchange constant of $J_{ij}=J$ between magnetic moments
$i$ and $j$.
The second term is the interfacial Dzyaloshinskii–Moriya
interaction, where $\mathbf{D}_{ij}=D\mathbf{\hat{z}}\times\mathbf{\hat{r}}_{ij}$ is the Dzyaloshinskii–Moriya
vector between magnetic moments $i$ and $j$ and $\mathbf{\hat{r}}_{ij}$
is the unit distance vector between sites $i$ and $j$.
Here, $\langle i, j\rangle$ indicates that the sum is performed only
over the nearest neighbors of the $i$th magnetic moment.
The third term is the Zeeman interaction with an applied external magnetic
field $\mathbf{H}$,
where $\mu=g\mu_B$ is the magnitude of the atomic magnetic moment, $g=|g_e|=2.002$ is the electron
$g$-factor, and $\mu_B=9.27\times10^{-24}$ J T$^{-1}$ is the
Bohr magneton. The fourth term is the 
sample anisotropy, with anisotropy strength $K_0$, and the
last term is the defect anisotropy with anisotropy strength $K_d$.
The defects are contained in the set $N$, and are modeled as
randomly located higher anisotropy lattice sites.
Defects are placed only within
periodic stripes in the sample, as
illustrated in Fig~\ref{fig:1}.
The long-range dipolar interaction acts as an
anisotropy in ultrathin films (see Supplemental
Material of Wang {\it et al.}\cite{wang_theory_2018}),
and therefore the effect of including
such an interaction is to shift the anisotropy
values.

The time evolution of the atomic magnetic moments 
is obtained using the LLG
equation \cite{seki_skyrmions_2016, gilbert_phenomenological_2004}:

\begin{eqnarray}\label{eq:2}
  \partial_t\mathbf{m}_i=&-\gamma\mathbf{m}_i\times\mathbf{H}^\mathrm{eff}_i
  +\alpha\mathbf{m}_i\times\partial_t\mathbf{m}_i\\\nonumber
  &+\frac{j\hbar\gamma P a^2}{2e\mu}\mathbf{m}\times\left(\hat{\mathbf{j}}\times\hat{\mathbf{z}}\right)\times\mathbf{m} \ .
\end{eqnarray}
Here $\gamma=1.76\times10^{11}$ T$^{-1}$ s$^{-1}$ is the electron gyromagnetic ratio,
$\mathbf{H}^\mathrm{eff}_i=-\frac{1}{\mu}\frac{\partial \mathcal{H}}{\partial \mathbf{m}_i}$
is the effective magnetic field including all interactions from
the Hamiltonian, $\alpha$ is the phenomenological damping
introduced by Gilbert, and the last term is the
torque induced by the spin Hall effect, where $j$
is the current density, $P=1$ is the spin polarization,
$e$ is the electron charge,
and $\hat{\mathbf{j}}=\hat{\mathbf{x}}$ is the current direction.

We fix the magnetic field value
$\mu\mathbf{H}=0.5(D^2/J)\mathbf{\hat{z}}$
in our simulations.
The material parameters are $J=1$ meV, $D=0.2J$,
$K_0=0.01J$, $K_d=0.02J$, and $\alpha=0.3$.
For each simulation, the system is initialized with a skyrmionium
in a clean region at the left edge of the sample, as shown
in Fig~\ref{fig:1}.
The numerical integration of Eq.~\ref{eq:2} is performed using
a fourth order Runge-Kutta method over 200 ns.

In Fig.~\ref{fig:1}, we show a system in which
the skyrmionium has been initialized 
on the left side of the sample.
Along the $x$ direction,
the sample alternates between
disorder free and pinned regions. In this work, we hold the total area of the
pinned regions constant and
vary the density of the defects in the pinned regions,
giving a defect density $\rho_{\mathrm{def}}=N_{\rm pin}/N_{\rm spins}$,
where $N_{\rm spins}$ is the total number of spins in the sample and
$N_{\rm pin}$ is the total number of defected spin sites.
For the system in Fig.~\ref{fig:1}, $\rho_\mathrm{def}=17\%$.
The skyrmionium is initially placed in a clean region, and the applied current drives the skyrmionium in the positive $ x$ direction.

\section{Results}

\begin{figure}
  \centering
  \includegraphics[width=0.7\columnwidth]{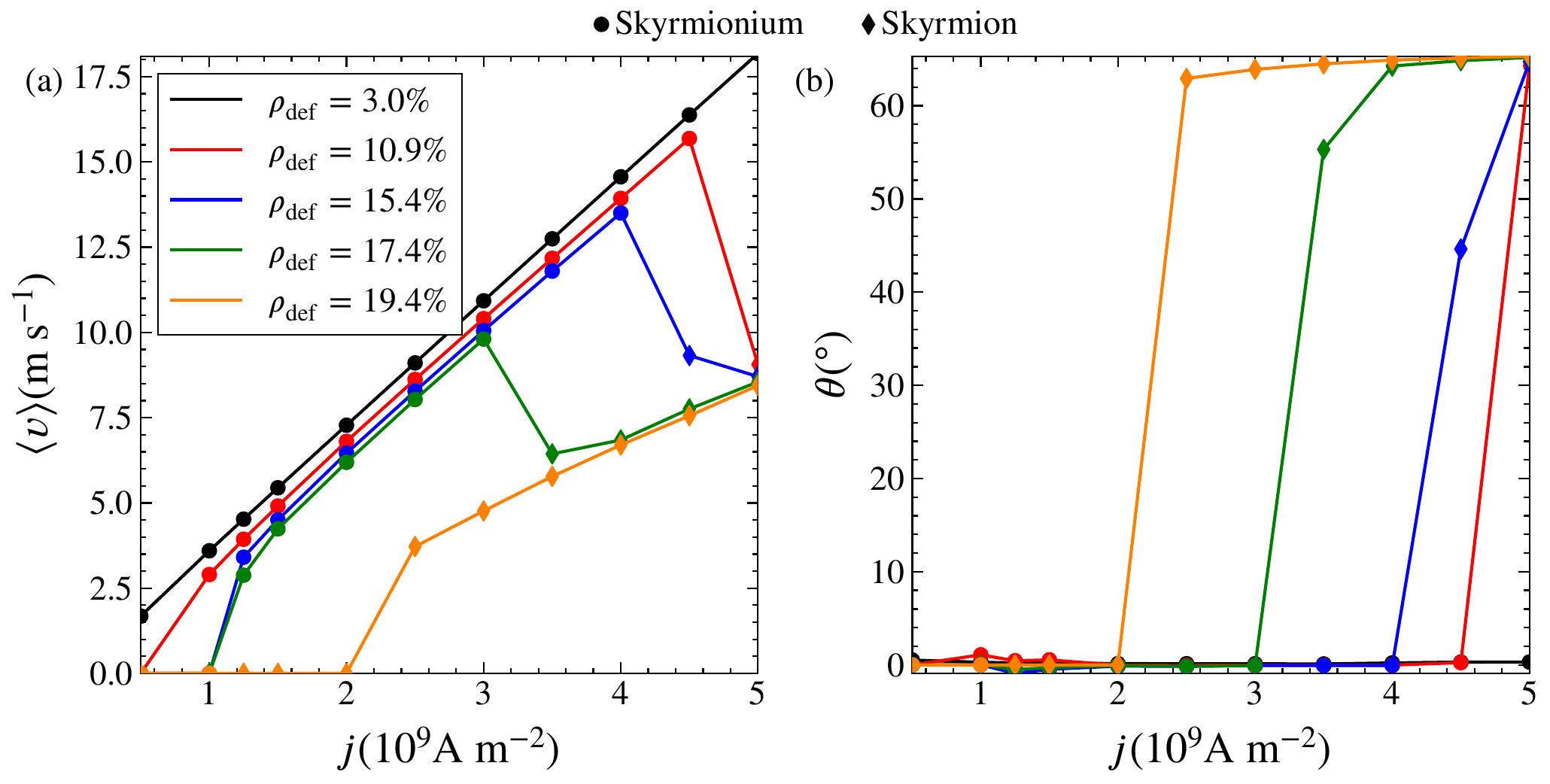}
  \caption{(a) Average velocity $\left\langle v\right\rangle$ vs $j$ and
    (b) the corresponding Hall angle $\theta$ vs $j$ for
    different defect densities $\rho_\mathrm{def}=3.0\%$ (black), $10.9\%$ (red),
    $15.4\%$ (blue), $17.4\%$ (green), and $19.4\%$ (orange).
    Circles indicate a moving skyrmionium, which has a zero Hall angle,
    and diamonds are for a moving skyrmion.}
  \label{fig:2}
\end{figure}

In Fig.~\ref{fig:2}, we plot the average velocity of the texture $\left\langle v\right\rangle$ vs $j$ for the system in Fig.~\ref{fig:1} at varied
$\rho_\mathrm{def}=3.0\%$,
$10.9\%$,
$15.4\%$,
$17.4\%$,
and $19.4\%$.
For $\rho_\mathrm{def}=3.0\%$, the velocity increases linearly with $j$, and the texture remains a
skyrmionium for drives all the way up to
$5\times 10^9$ A m$^{-2}$.
There is a
higher critical current (not shown) where the skyrmionium becomes unstable
in the clean portion of the sample.
For $\rho_{\mathrm{def}}=10.9\%$,
$15.4\%$,
and $17.4\%$,
we find a finite depinning threshold that increases with
increasing $\rho_{\mathrm{def}}$.
At a critical driving force $j_c$,
there is a drop down in the velocity, and above this current the
velocity resumes its linear increase with increasing $j$ from its
new lower value.
In Fig.~\ref{fig:2}(b), we plot the 
Hall angle of the texture $\theta$ versus $j$.
For currents smaller than $j_c$,
the Hall angle is zero,
$\theta = 0$, as expected for skyrmionium.
At $j_c$, 
the Hall angle jumps up from zero to
$\theta = 60^\circ$.
The drop in the velocity is associated with a transition
from a moving skyrmionium with $Q = 0$ and a zero Hall angle
to a moving skyrmion with $Q = -1$ and a finite Hall angle.
For $\rho_{\mathrm{def}}=19.4\%$,
there is an extended range of current over which
the skyrmionium remains pinned, and once the texture begins to move, it
immediately transforms into a skyrmion,
indicating that
there is a critical defect density
above which a moving skyrmionium is not stable.

\begin{figure}
  \centering
  \includegraphics[width=\columnwidth]{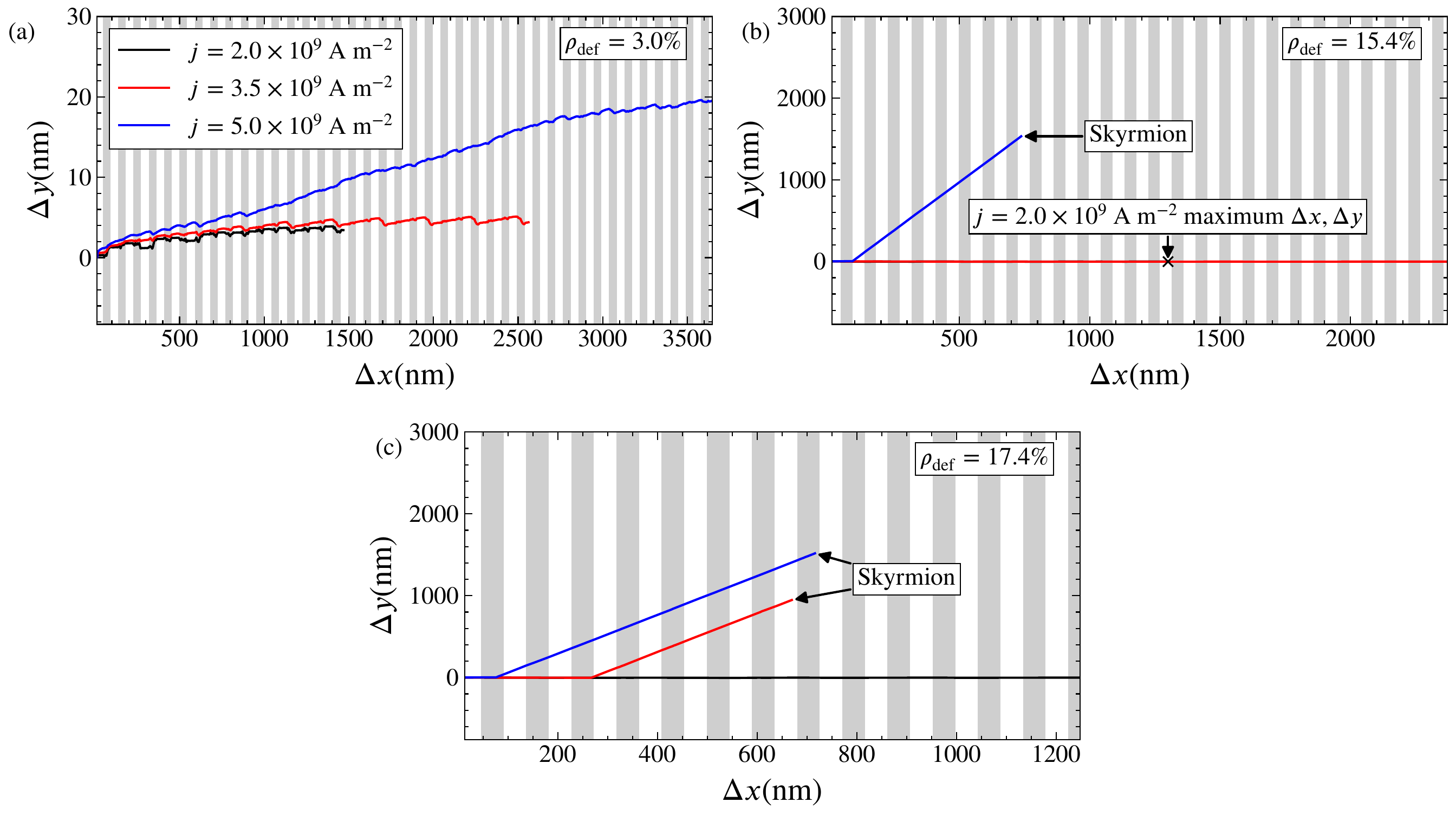}
  \caption{Net cumulative displacements $\Delta x$ and $\Delta y$
    for different currents of $j=2\times10^9$ A m$^{-2}$ (black),
    $j=3.5\times10^9$ A m$^{-2}$ (red), and $j=5\times10^9$ A m$^{-2}$ (blue)
    at different defect densities of
    (a) $\rho_\mathrm{def}=3.0\%$,
    (b) $\rho_\mathrm{def}=15.4\%$, and
    (c) $\rho_\mathrm{def}=17.4\%$.
    In each case the total elapsed time is the same.
    The gray stripes represent regions with defects, whereas white
    stripes represent clean regions.
    The sample is of the size shown in Fig.~\ref{fig:1} but has been
    expanded out along the periodic boundary conditions in these images.
    In some cases, the skyrmionium transforms into a skyrmion
    with a finite skyrmion Hall angle after a
    transient period of motion.
    Animations showing the skyrmionium and/or skyrmion motion
    appear in the Supplemental material \cite{supplemental}.}
  \label{fig:3}
\end{figure}

In Fig.~\ref{fig:3}, we show the trajectory of the textures,
where we plot the displacements $\Delta x$ and $\Delta y$
over a fixed total time for
different currents of $j=2\times10^9$ A m$^{-2}$,
$j=3.5\times10^9$ A m$^{-2}$,
and $j=5\times10^9$ A m$^{-2}$
at different defect densities.
At $\rho_\mathrm{def}=3.0\%$ in
Fig.~\ref{fig:3}(a),
the texture remains a skyrmionium for all of the drives we consider.
This skyrmionium travels
primarily along the $x$ direction, and only makes a small excursion
into the $y$ direction for
$j=5\times10^9$ A m$^{-2}$, where it moves
nearly $3500$ nm along the driving direction.
In Fig.~\ref{fig:3}(b) at $\rho_\mathrm{def}=15.4\%$,
the skyrmionium moves only along $x$
for $j=2\times10^9$ A m$^{-2}$ and
$j=3.5\times10^9$  A m$^{-2}$,
traveling a distance of nearly 2500 nm in the latter case.
When $j=5\times10^9$ A m$^{-2}$,
the skyrmionium transitions into a skyrmion after moving a distance of $100$ nm
in the $x$ direction. After this, it moves with a finite Hall angle
and translates approximately $1600$ nm along the $y$ direction and $750$ nm
along the $x$ direction.
Here, the skyrmion moves more slowly and also at an angle to the drive
compared to the skyrmionium.
For $\rho_\mathrm{def}=17.4\%$ in
Fig.~\ref{fig:3}(c),
at the lowest current of $j=2\times10^9$ A m$^{-2}$ the
texture remains a skyrmionium and moves in the $x$ direction
a total of $1200$ nm.
This indicates that
as the density of the disorder increases,
the distance the skyrmionium can travel during a fixed time decreases.
For $j=3.5\times10^9$ A m$^{-2}$ and $j=5\times10^9$ A m$^{-2}$, the
texture transitions into a skyrmion
and moves at an angle to the drive.
The rapidity of this transformation depends on the drive.
For
$j=3.5\times10^9$ A m$^{-2}$,
the skyrmionium travels $250$ nm before transforming into a skyrmion,
while for $j=5\times10^9$ A m$^{-2}$,
it travels only $70$ nm.
This indicates that there is a time dependent process that occurs
during the skyrmionium transformation that is affected by
the disorder density and the drive.

\begin{figure}
  \centering
  \includegraphics[width=\columnwidth]{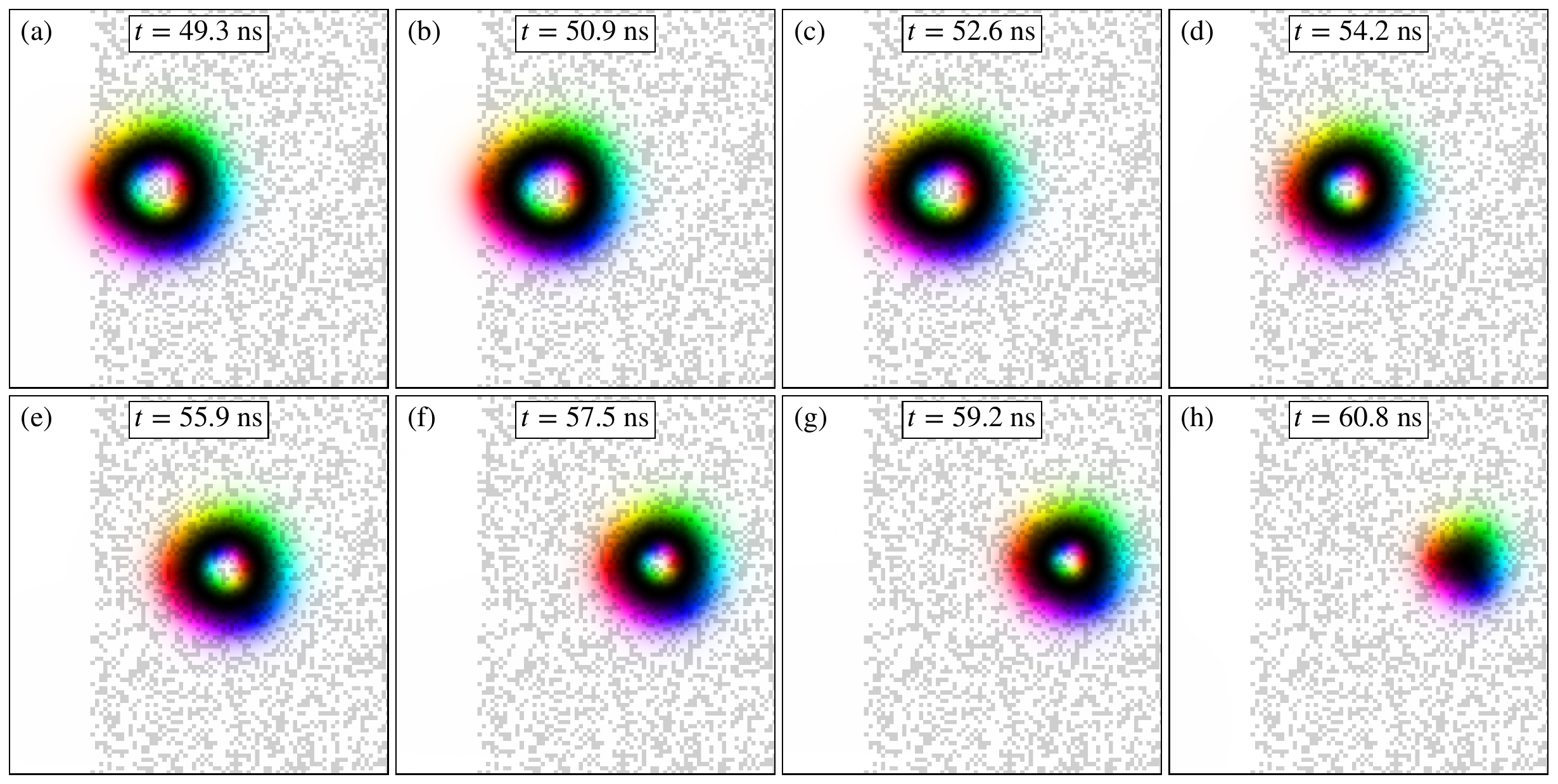}
  \caption{Images illustrating the transformation of a skyrmionium into a 
    skyrmion as a result of interactions with defects in a sample with
    $j=1\times10^9$ A m$^{-2}$ and
    $\rho_\mathrm{def}=17.4\%$.
    The inner skyrmion shrinks until it collapses.
    An animation showing the skyrmionium to skyrmion
    transformation appears in the Supplemental Material
    \cite{supplemental}.}
  \label{fig:4}
\end{figure}

In Fig.~\ref{fig:4}, we illustrate the
time dependent transformation of a skyrmionium into a
skyrmion as it enters the pinned region
for a system with $j=1\times10^9$ A m$^{-2}$ and
$\rho_\mathrm{def}=17.4\%$.
Here, the inner skyrmion shrinks until it collapses,
similar to the behavior observed by Zhang {\it et al.}\cite{zhang_control_2016}
upon increasing an external magnetic field. After the collapse, the
skyrmion moves away from the collapse point with
a finite skyrmion Hall angle. 

\begin{figure}
  \centering
  \includegraphics[width=\columnwidth]{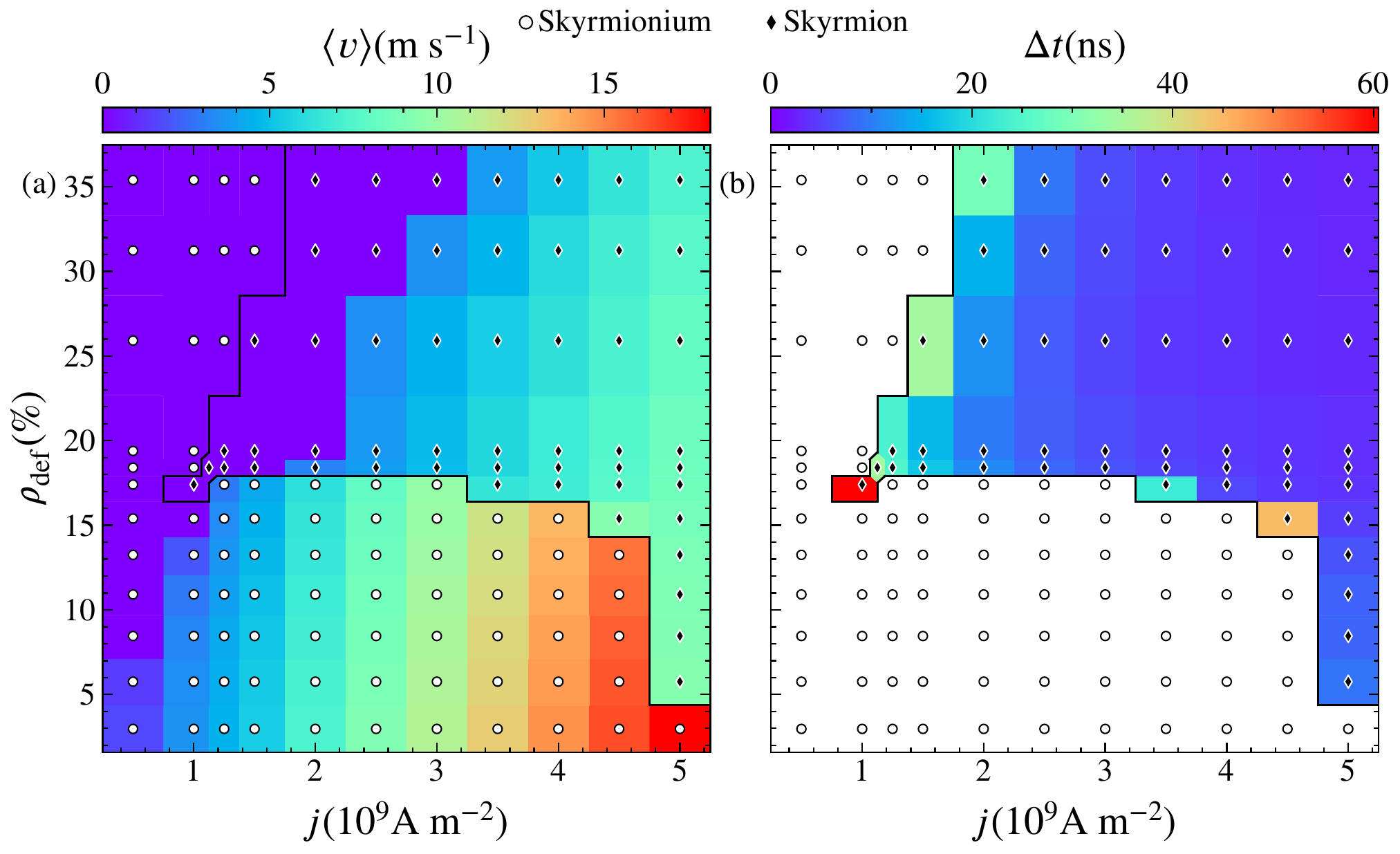}
  \caption{Heat diagrams of (a) average velocity
    $\left\langle v\right\rangle$ and (b) time required
    for the skyrmionium to transform itself into a skyrmion $\Delta t$
    plotted as a function of $\rho_\mathrm{def}$ vs
    $j$. Circles indicate skyrmionium, and diamonds
    indicate skyrmions.
    The black solid line is the border between the
    skyrmionium regime and the skyrmion regime in both panels,
    and the heavy black line in panel (a) separates the pinned
    skyrmionium state with $\langle v\rangle=0$ from the
    flowing states.
    In the white regions in (b), the skyrmionium
    never transforms into a skyrmion, giving $\Delta t\to\infty$.}
  \label{fig:5}
\end{figure}

In Fig.~\ref{fig:5}(a), we plot a heat diagram of
the average velocity
$\left\langle v\right\rangle$ as a function of $\rho_\mathrm{def}$
versus $j$
as measured from velocity-force curves.
We observe
a pinned phase,
moving skyrmionium, and
moving skyrmions.
The solid lines indicate the
boundaries between the different states.
For  $\rho_\mathrm{def} < 17\% $ we find both pinned and moving
skyrmionium states.
The velocity in the moving skyrmionium regime
decreases with increasing $\rho_\mathrm{def}$ for fixed $j$.
In the same window of $\rho_\mathrm{def} < 17\% $,
there is a critical driving force $j_c$
above which the skyrmionium transforms into
a skyrmion. This drive is accompanied by a drop in the velocity of the texture.
The critical drive $j_c$
decreases with increasing $\rho_\mathrm{def}$.
When $\rho_\mathrm{def} > 17\%$, we find only
pinned skyrmionium and moving skyrmion states since
the skyrmionium transforms into a skyrmion upon depinning,
indicating that there is a critical disorder strength
above which a moving skyrmionium is unstable.

In Fig.~\ref{fig:5}(b) we show a heat map of the time required
for the skyrmionium to transform into a skyrmion, $\Delta t$,
as a function of $\rho_{\mathrm{def}}$ versus $j$.
Here, white regions indicate points at which
the skyrmionium never transforms into a skyrmion.
In general, the transformation time is longest near the depinning threshold.
We find an interesting reentrance effect at
$\rho_\mathrm{def} = 15\% $,
where the skyrmionium transforms into a skyrmion upon depinning at
$j=1\times10^9$ A m$^{-2}$. This transformation appears
in Fig.~\ref{fig:5}(a), and is also associated with
the longest transient time in Fig.~\ref{fig:5}(b).
For this same defect density, over the interval
$1\times10^9~\mathrm{A m}^{-2} < j <  3\times10^9~\mathrm{A m}^{-2}$
a moving skyrmionium is stable, while
for $j > 3\times10^9$ A m$^{-2}$ the skyrmionium becomes unstable again.
The reentrance appears because there are two effects that can
destabilize the skyrmionium.
The first is the quenched disorder, which
can induce a destabilizing roughness of the skyrmionium boundary,
particularly near the depinning transition. The second is the current itself,
which distorts the skyrmionium.
From previous studies of driven particles moving over quenched disorder,
it is known that the effective roughening produced by the quenched
disorder can be diminished
in the moving state
at higher drives \cite{Reichhardt15,Reichhardt17},
suggesting that the drive can actually
reduce some of the roughness induced in the skyrmionium boundary by
the pinning when the skyrmionium is in a moving state.
If this reduction is greater than the roughness
generated by the current itself,
an applied drive can stabilize the
moving skyrmionium
over a certain range of quenched disorder densities and currents.
Near the critical disorder density,
there can be a window in which
the drive decreases the distortion of the
moving skyrmionium created by the pinning, and yet where the drive
remains low enough that the current does not strongly distort the
skyrmionium. The extent of this reentrance
depends on the size scale of the disorder and other parameters.

\begin{figure}
  \centering
  \includegraphics[width=0.5\columnwidth]{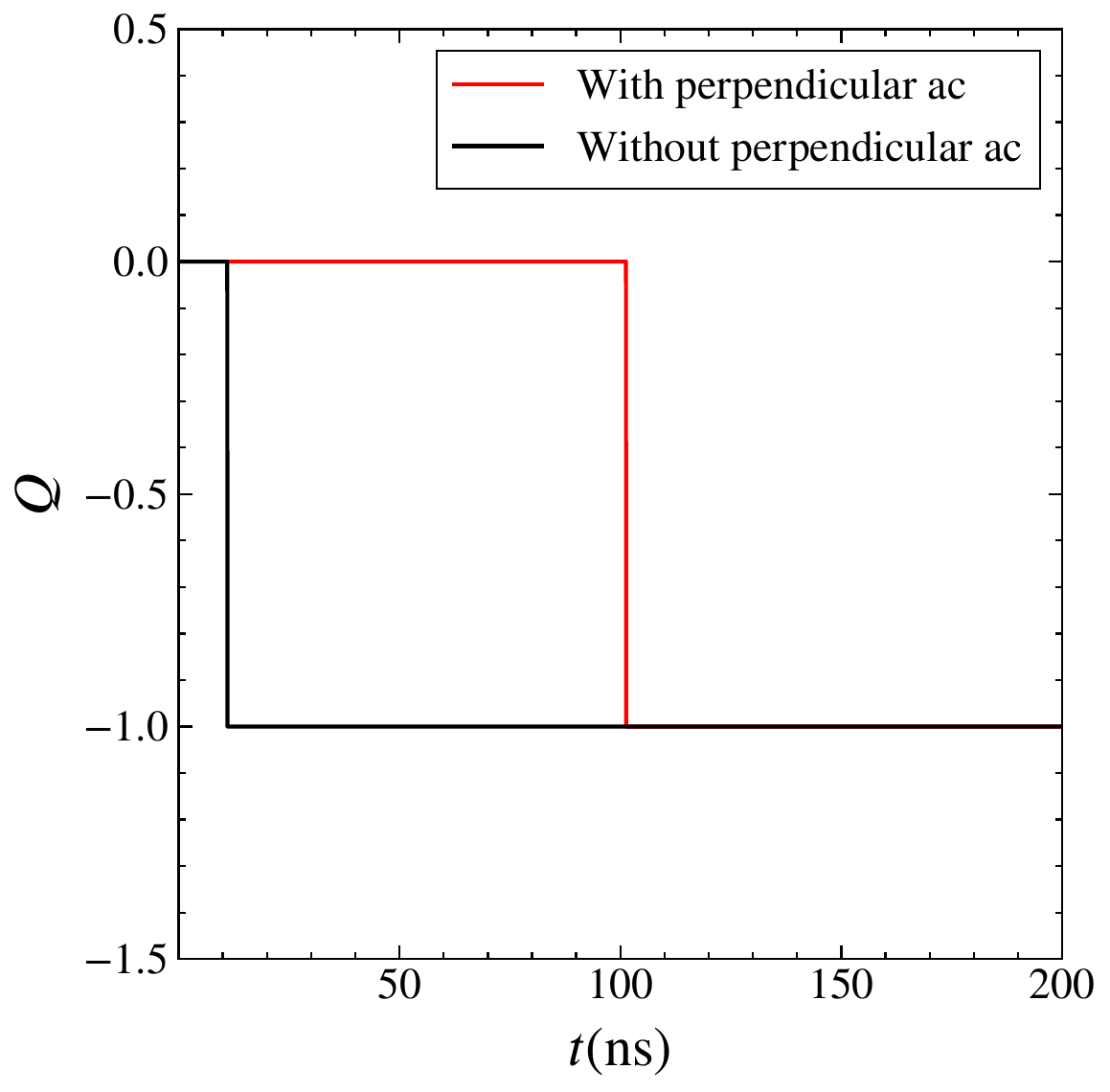}
  \caption{ 
    The topological charge $Q$ versus time for a system with $j=2\times10^9$ A m$^{-2}$ and $\rho=17.4$\%.
The black line is for a system with only a dc drive applied along the $x$
 direction.
The red line is for the same dc drive along with
an ac drive applied in the $y$ direction,
${\bf j}_\mathrm{ac}=j_\mathrm{ac}\sin(2\pi\omega t){\bf \hat{y}}$, with $j_\mathrm{ac}=2\times10^9$ A m$^{-2}$ and $\omega=0.07$ GHz.
Without perpendicular ac driving,
the skyrmionium survives for approximately 10 ns,
but with perpendicular ac driving, it survives for approximately 100 ns.}
  \label{fig:6}
\end{figure}

To test whether a transverse drive can reduce the effects of the
quenched disorder,
we applying
a small transverse ac shaking drive
to the dc driven skyrmionium. This causes the skyrmionium to move
a small amount back and forth in
the $y$ direction in addition to translating along $x$.
Above the critical disorder density, we find that the additional
ac drive can increase the lifetime of
the skyrmionium, as shown in Fig.~\ref{fig:6}
where we plot the topological charge $Q$ versus time for a system
with $j=2\times10^9$ A m$^{-2}$, $\rho_\mathrm{def}=18\%$, and
a transverse ac drive of
${\bf j}_\mathrm{ac}=j_\mathrm{ac}\sin(2\pi\omega t){\bf \hat{y}}$,
where $j_\mathrm{ac}=2\times10^9$ A m$^{-2}$ and $\omega=0.07$GHz.
When only a dc drive is applied,
the $Q=0$ 
skyrmionium transforms to a
$Q=-1$ skyrmion in less than $10$ ns.
When transverse ac driving is added,
the skyrmionium remains stable for $100$ ns,
an increase of ten times
compared to the purely dc driving situation.
We have also considered other
ac driving amplitudes and frequencies,
and find that the stabilizing enhancement
is the most prominent when
$j_\mathrm{ac} < 4\times10^9$ A m$^{-2}$, since
overly large
ac amplitude driving induces extra distortions that destabilize the
skyrmionium instead of stabilizing it.

We expect the general features in the heat diagrams
of Fig.~\ref{fig:5} to remain robust
for uniform disorder or other types of disorder patterns beyond
the periodic array considered here,
which would limit
the possible
regimes in which
a moving skyrmionium can remain stable.
We only considered a single skyrmionium; however,
it is possible that a skyrmionium lattice might remain
stable up to higher drives if the interactions between the textures
are able to
reduce the distortions that occur near the depinning transition.
Previous work has shown that
increasing the elastic interactions between neighboring particles
can reduce the pinning threshold
produced by quenched disorder \cite{Reichhardt17}.
It would also be interesting to study thermal
effects. These could, on their own,
destabilize the skyrmionium; however, thermal effects can also reduce the
effectiveness of the quenched disorder,
so it may be possible to observe
additional reentrant phases at finite temperatures.

\section{Summary}

We have used atomistic simulations to investigate the driven dynamics of skyrmionium moving over a periodic disorder array for varied defect densities and drives.
For low defect densities, the skyrmionium engages in stable motion
for an extended range of currents.
As the disorder density increases, there is a critical driving force
above which the skyrmionium transforms into a skyrmion.
This transition is accompanied by a sudden drop in the
velocity at the critical driving force $j_c$
along with the onset of a finite skyrmion Hall angle.
The value of $j_c$
decreases with increasing disorder density until a
critical disorder density
is reached above which a moving skyrmionium is unstable and
the skyrmionium transitions to a skyrmion at the depinning transition.
We map out a dynamic phase diagram as a function of disorder density
versus current and find a pinned phase,
a moving skyrmionium state, and a moving skyrmion state.
Near the critical disorder strength,
we find a reentrant behavior in which a drive just above depinning
destabilizes the skyrmionium into a skyrmion,
but an intermediate drive can stabilize the moving skyrmionium.
At high drives, the moving skyrmionium
again transitions into a moving skyrmion.
This effect results when a drive in the reentrant window is able to
reduce the roughening effect of the quenched disorder on the walls of the
skyrmionium.
We find that adding a transverse ac drive can enhance the stability
of a moving skyrmionium since the ac drive acts as a shaking term
that reduces the effect of the pinning on the texture.
Our results indicate that while skyrmionium moves faster
than a skyrmion in the presence of quenched disorder,
the skyrmionium only remains stable
over a limited range of current values and quenched disorder densities.

\section*{Acknowledgments}
This work was supported by the US Department of Energy through the Los Alamos National Laboratory. Los
Alamos National Laboratory is operated by Triad National Security, LLC, for the National Nuclear Security
Administration of the U. S. Department of Energy (Contract No. 892333218NCA000001). 
J.C.B.S  and N.P.V. acknowledge funding from Fundação de Amparo à Pesquisa do Estado de São Paulo - FAPESP (Grants J.C.B.S 2023/17545-1 and 2022/14053-8, N.P.V 2024/13248-5).
We would like to thank FAPESP for providing the computational resources used in this work (Grant: 2024/02941-1). 

\section*{References}
\bibliographystyle{iopart-num}
\bibliography{mybib}

\end{document}